\documentclass[12pt]{article}
\usepackage{amssymb,amsmath,epsfig,cite}
\usepackage{graphicx}
\usepackage{subfig}
\begin{document}
\title{\bf Energy Conditions in Higher Derivative $f(R,\Box R,T)$ Gravity}

\author{Z. Yousaf$^1$ \thanks{zeeshan.math@pu.edu.pk}, M. Sharif$^1$ \thanks{msharif.math@pu.edu.pk}
M. Ilyas$^2$ \thanks{ilyas\_mia@yahoo.com} and M. Z. Bhatti$^1$ \thanks{mzaeem.math@pu.edu.pk}\\
$^1$ Department of Mathematics, University of the Punjab,\\
Quaid-i-Azam Campus, Lahore-54590, Pakistan\\
$^2$ Centre for High Energy Physics, University of the Punjab,\\
Quaid-i-Azam Campus, Lahore-54590, Pakistan}

\date{}

\maketitle
\begin{abstract}
In this paper, we examined the viability bounds of a higher derivative $f(R,\Box R, T)$
theory through analyzing energy conditions (where $R,~\Box R$ and
$T$ are the Ricci scalar, d'Alembert's operator and trace of energy
momentum tensor, respectively). We take flat
Friedmann-Lema\^{i}tre-Robertson-Walker spacetime coupled with ideal
configurations of matter content. We consider three different
realistic models of this gravity, that could be utilized to
understand the stability of cosmological solutions. After
constructing certain bounds mediated by energy conditions, more
specifically weak energy condition, we discuss viable zones of the
under considered modified models in an environment of recent
estimated numerical choices of the cosmic parameters.
\end{abstract}
{\bf Keywords:} Relativistic fluids; Gravitation; Stability\\
{\bf PACS:} 04.50.Kd; 04.20.-q; 98.80.Jk; 98.80.-k\\

\section{Introduction}

The Einstein's theory of General Relativity (GR) is considered as a
fundamental theory to understand several hidden aspects of
gravitational dynamics that has produced many viable results in
outlining our universe picture. The cosmological data stem from,
e.g., the Wilkinson Microwave anisotropy probe
(WMAP)~\cite{Komatsu:2010fb, Hinshaw:2012aka}, the BICEP2
experiment~\cite{Ade:2014xna, Ade:2015tva, Array:2015xqh} and the
Planck satellite~\cite{ya2, Planck:2015xua, Ade:2015lrj},, point
that our cosmos is expanding with an accelerating rate. The outcomes
from Planck satellite illustrate dark energy (DE), the enigmatic
force, as a dominant constituent among all other ingredients of our
cosmos. Various techniques have been proposed in order to comprehend
this dominant mysterious element. Qadir et al. \cite{zs3} examined
various important feature of modified relativistic dynamics and
recommended that GR may need to modify to resolve issues related to
quantum gravity and dark matter (DM) problem. Some well-motivated
modified gravity theories (MGTs) have been suggested by various
notable physicists and gained much interest. The MGTs are based on
the approach of plugging some generic curvature invariant functions,
instead of Ricci scalar in the formula of the Einstein-Hilbert (EH)
action with (for further reviews on DE and MGTs, see, for
instance,~\cite{R1,R2,R3,R4,R5,R6,R7,R8,R9,R10,3,4,5,6,8}).

Nojiri and Odintsov \cite{no1} introduced first observationally
well-consistent accelerating cosmic model through $f(R)$ gravity.
The same authors \cite{bin4} proposed that some MGTs could provide a
platform that can evidently describe various cosmological scenario.
The effects of dark source terms coming from Einstein-$\Lambda$
\cite{zs1}, $f(R)$ \cite{z1fr}, $f(R,T)$ \cite{z2frt} ($T$ is the
trace of energy momentum tensor) and $f(R,T,R_{\mu\nu}T^{\mu\nu})$
gravity \cite{z3frtrmn}, on some dynamical features of
self-gravitating stellar systems have been discussed recently. The
simplest modification of GR is $f(R)$ theory, in which one has used
arbitrary Ricci scalar function instead of the Ricci scalar in the
EH action. The model of $f(R)$ gravitational theory was proposed,
after couple of years from the arrival of GR to consider modified
relativistic dynamics \cite{ya4}. Afterwards, this toy model was
examined intermittently by various people \cite{ya5} with the aim to
re-normalize GR \cite{ya6}.

In this direction, several MGTs have been introduced, like, $f(G)$
(where $G$ is the Gauss-Bonnet term), $f(R,G)$ and
$f(R,R_{\mu\nu}^{~~\mu\nu},R_{\mu\nu\lambda\gamma}^{~~~~\mu\nu\lambda\gamma})$.
Harko \textit{et al.} \cite{ya9} generalized $f(R)$ gravity to
$f(R,T)$ theory. The inclusion of $T$ corrections is based on the
physical grounds of introducing quantum effects or exotic
relativistic fluid configurations. Based on the concept of matter
geometry coupling, they presented field as well as dynamical
equations of motion with some particular $f(R,T)$ models. Recently,
Houndjo et al. \cite{bx1} put forward the basic concept of $f(R,T)$
by introducing $\Box R$ term in $f(R,T)$ action. This term often
appears in studying some dynamical features of string theory, which
was firstly considered by \cite{bx2}. Such kind of gravitational
models could give gravitational results compatible with scalar field
theories under some limits. If someone uses generic
higher-derivative mathematical term in modified action, then after
making conformal transformations, that toy model could provide
dynamical background for canonical Einstein theory along with finite
scalar degrees of freedom \cite{bx3}. Sharif and his collaborators
probed the impact of various realistic formulations of modified
gravity models on the evolutionary phases of compact structures
\cite{ps2} as well as cosmology \cite{ps3}.

It could be possible that any arbitrary configuration of matter
field satisfy field equations. But that formulations may not
correspond to any physical environment. In order to represent
realistic form of this tensorial quantity, it should obey certain
constraints. These conditions are coordinate-invariant that stem
from the well-known Raychaudhuri equations \cite{21,22,23}, known as
energy condition (ECs). These ECs are of four types namely, weak
energy conditions (WEC), null energy conditions (NEC), strong energy
conditions (SEC) and dominant energy conditions (DEC). Visser
\cite{bx4} studied the phenomenon of galaxy formation with the help
of ECs and proposed that during the evolutionary transformation of
the current and galaxy forming eras, the matter must be subjected to
disobey SEC. Santos et al. \cite{bx5} described the peculiar
relations between ECs and cosmological observations by obtaining a
model-independent constraints on the luminosity distance of the
energy distributions of our flat expanding universe.

Santos et al. \cite{bx6} applied the concept of ECs to put viability
constraints on $f(R)$ models and noticed that $f(R)=R+\alpha R^2$
(with $\alpha>0$) violates WEC. Santos et al. \cite{bx7} studied the
cosmic accelerating phases through ECs and pointed out that three of
the ECs, related with phantom fields, must be disobeyed under
certain constraints. Bertolami and Sequeira \cite{bx8} used the
notion of ECs to present a somehow general form of viable $f(R)$
model and claimed that such model could be used to unveil various
interesting features of this gravity. Atazadeh et al. \cite{bx9}
used the concept of $f(R)$ ECs and applied these in Brans-Dick
theory to get some physically acceptable results about the expanding
cosmos.

Sen and Scherrer \cite{bx10} examined the role of ECs on the DM and
DE contents of the flat universe models and evaluated some bounds on
the Hubble parameter in order to study the expanding behavior of our
cosmos. Nojiri \emph{et al.} \cite{15} evaluated viability bounds
through ECs in order to present observationally viable $f(G)$ model.
Bamba et al. \cite{bx11} determined some viability bounds for the
physical applicability of $f(G)$ models by using current values of
the Hubble parameter. In the recent times, the conception of ECs in 
the examinations of the realistic cosmic models has been performed for Einstein-$\Lambda$ 
\cite{bx12}, $f(G)$  \cite{bani1, sad1}, $f(R,G)$ \cite{bx13} gravity. The theoretical formations of wormholes
structures has fascinated many researchers in the literature \cite{ov1,ov2,ov3}, while
their connection with ECs has also been studied \cite{til2,til3,til1}.

In this work, we have probed the problem of viability of some modified
gravity models. We have chosen recent choices of the jerk, cackle, deceleration,
snap as well as Hubble parameters. After employing certain bounds
induced from $f(R,\Box R, T)$ gravity, we discuss ECs that could help to indicate
certain viability areas in the model building. The present work is organized
as follows. In the coming section, we introduce briefly $f(R,\Box R, T)$ equations
of motion, while modified mathematical expressions of of ECs are presented
in the section 3. Section 4 is devoted to examine viability regimes of modified
ECs with three different realistic $f(R,\Box R, T)$ models. In the last section, we
summarize our main findings.

\section{$f(R,\Box R, T)$ Gravity}

The modified version of EH action for $f(R,\Box R, T)$ gravity can be given as
\begin{equation}\label{action}
S = \frac{1}{2\kappa^2}\int {{d^4}x\sqrt { - g} f(R,\Box R, T) + {S_M}\left( {{g^{\mu \nu }},\psi } \right)},
\end{equation}
where $\kappa$ is the coupling constant with ${\kappa ^2} = 8\pi G$,
$R$ and $T$ are the traces of the Ricci and usual energy-momentum
tensors, respectively. Furthermore, $\Box$ is the de'Alembert
operator that can be expressed through covariant derivative,
$\nabla_\mu$, as $\Box\equiv\nabla_\mu \nabla^\mu$. By giving
variations in the above equation with respect to metric tensor, we
have
\begin{equation}\label{action1}
\delta S = \frac{1}{2\kappa^2}\int {d^4}x[f\delta \sqrt{-g}+\sqrt{-g}(f_R\delta R+f_{\Box R}\delta\Box R+f_T\delta T)
+2\kappa^2 \delta(\sqrt{-g}L_M)],
\end{equation}
where $L_M$ is the matter Lagrangian. Using the values of $\delta
R,~\delta\Box R$ and $\delta T$ and $\delta \sqrt{-g}$ in the above
equation, we obtain an equation, which after some manipulations,
provides
\begin{align}\nonumber
\delta S &= \frac{1}{2\kappa^2}\int {d^4}x\left[-\frac{1}{2}\sqrt{-g}g_{\alpha\beta}\delta g^{\alpha\beta}f
+\sqrt{-g}(T_{\alpha\beta}+\Theta_{\alpha\beta})f_T\delta g^{\alpha\beta}
+f_R\sqrt{-g}(R_{\alpha\beta}\right.\\\nonumber
&\left.+g_{\alpha\beta}\Box-\nabla_\alpha\nabla_\beta)\delta g^{\alpha\beta}
+\sqrt{-g}f_{\Box R}(\nabla_\alpha\nabla_\beta R+\Box R_{\alpha\beta}+g_{\alpha\beta}\Box^2
+R_{\alpha\beta}\Box-\Box \right.\\\label{action2}
&\left.\times\nabla_\alpha\nabla_\beta-\nabla_\alpha R\nabla_\beta+2g^{\mu\nu}\nabla_\mu
R_{\alpha\beta}\nabla_\nu)\delta g^{\alpha\beta}+2\kappa^2\frac{\delta(\sqrt{-g}L_M)}{\delta g^{\alpha\beta}}\delta
g^{\alpha\beta}\right],
\end{align}
where subscripts $R,~T$ and $\Box R$ indicate the derivative of the
corresponding quantities with respect to $T,~R$ and $\Box R$,
respectively, while $\Theta_{\alpha\beta}=g^{\mu\nu}\delta
T_{\mu\nu}/\delta g^{\alpha\beta}$. Equation (\ref{action2}) after
simplifications gives rise to
\begin{equation}\label{fieldeq}
\begin{gathered}
{{f_R}{R_{\alpha\beta}}} + \left( {{g_{\alpha\beta}}\Box  - {\nabla _\alpha }{\nabla _\beta }} \right){{f_R}}
- \frac{1}{2}{g_{\alpha\beta}}f + \left( {2{f_{\Box R}}
({\nabla _{(\alpha }}{\nabla _{\beta )}})} \right.R\left. { - \Box {R_{\alpha\beta}}} \right)\\
- \left\{ {{R_{\alpha\beta}}}\Box - \Box {\nabla _\alpha }{\nabla _\beta }
+ {g_{\alpha\beta}}{\Box ^2} - {\nabla _\alpha }R{\nabla _\beta }
+2 {g^{\mu\nu }}{\nabla _\mu } {{R_{\alpha\beta}}{\nabla _\nu }} \right\}{f_{\Box R}}\\
=\kappa^2 T_{\alpha\beta}-f_T (T_{\alpha\beta}+\Theta_{\alpha\beta}).
\end{gathered}
\end{equation}

We consider our relativistic system by considering a
Friedmann-Lema\^{i}tre-Robertson-Walker (FLRW) metric
\begin{equation}\label{metric}
d{s^2} =  - d{t^2} + {a^2}(t)(d{x^2} + d{y^2} + d{z^2}),
\end{equation}
where $a$ indicates the scale factor. We further assume that
our geometry is coupled with the following perfect fluid
\begin{equation}\label{emtensor}
T_{\alpha\beta}=(\rho+p)u_\alpha u_\beta -p g_{\alpha\beta},
\end{equation}
where $u_\alpha$ is the fluid four vector, $\rho$ and $p$ are the
fluid's energy density and pressure, respectively. After using
matter Lagrangian to be $L_m=-p$, the value of the tensor
$\Theta_{\alpha\beta}$ has been found as ${\Theta_{\alpha\beta}=2
T_{\alpha\beta }-p g_{\alpha\beta}}$. With this background, the
$f(R,\Box R, T)$ field equations (\ref{fieldeq}) for the FLRW model
(\ref{metric}) filled with isotropic matter distributions
(\ref{emtensor}) provide
\begin{align}\nonumber
&2H{f_{\Box R}}^{\prime \prime \prime} - \left( {2{H^2} + 3H'}
\right){f_{\Box R}}^{\prime \prime } - (5{H^3} + 2HH'
+ {H^{\prime \prime }}){f_{\Box R}}^\prime + 2\{-2{H^2}H' + 6{{H'}^2} \\\label{tf1}
& + 3H{H^{\prime \prime }} + H^{\prime \prime \prime } \}
{f_{\Box R}} + H{f_R}^\prime - \left( {{H^2} + H'} \right){f_R}
- \frac{f}{6} = \frac{1}{3}[\rho {\kappa ^2} + (\rho  + p){f_T}],\\\nonumber
&{f_{\Box R}}^{\prime \prime \prime \prime} + 5H{f_{\Box R}}^{\prime \prime \prime}
+ \left( { - 8{H^2} + 5H'} \right){f_{\Box R}}^{\prime \prime } + ( - 23{H^3} + 2HH'
+ 4{H^{\prime \prime }}){f_{\Box  R}}^\prime \\\nonumber
&+ 2\left( { - 2{H^2}H' + 6{{H'}^2} + 3H{H^{\prime \prime }} + {H^{\prime \prime \prime }}} \right){f_{\Box R}}
 - 2H{f_R}^\prime - \left( {3{H^2} + H'} \right){f_R}\\\label{tf2}
& -{f_R}^{\prime \prime } - \frac{f}{2} = {\kappa ^2}p.
\end{align}
The parametric quantity corresponding to Hubble $(H)$ and deceleration $(q)$ for the case of FLRW model are
\begin{align}\label{hq}
H=\frac{\dot{a}}{a},\quad q =-\frac{1}{{{H^2}}}\frac{{a''}}{a},
\end{align}
whereas those parameters that correspond to jerk $(j)$, snap $(s)$ and crackle $(l)$ are found to be
\begin{align}\label{jsl}
j=\frac{{a'''}}{aH^3},\quad s = \frac{{a''''}}{aH^4}\quad
l=-\frac{{a'''''}}{aH^5}.
\end{align}

\section{Energy Conditions}

In the theoretical analysis of various interesting stellar
structures, like black holes, wormholes etc, the notion of energy
conditions occupies fundamental importance. The exploration about
the validity of these constraints has been a source of search engine
for many relativistic astrophysicists. It has been seen that stability regimes
of these ECs could assist enough to examine the stable arena of celestial structures.
It is well known that relativistic structures are coupled with matter configurations
which are described by their stress-energy tensors. In order to make the arbitrariness
of these tensors concise, it should mediate a realistic form of matter field. 
Those stress-energy tensors that obey ECs, could be
regarded as realistic ones. These conditions are
coordinate-invariant (independent of symmetry) restrictions on the
energy-momentum tensor. In the scenario of modified gravitational
theories, ECs can be obtained from the expansion nature of the
Raychaudhuri's equation given as follows
\begin{align}
\frac{d \Theta_1}{d\tau}=-\frac{\Theta_1}{2}+\omega^{\alpha\beta}\omega_{\alpha\beta}
-\sigma^{\alpha\beta}\sigma_{\alpha\beta}-R_{\alpha\beta}k^\alpha k^\beta,
\end{align}
where $\omega_{\alpha\beta},~\sigma_{\alpha\beta},~\Theta_1$ are
mathematical quantities describing rotation, shear and expansion
corresponding to congruences determined by null vector $kâ$.
Further, $R_{\alpha\beta}$ is the Ricci tensor. Sharif and his
collaborators \cite{ps1} have discussed viability regions for
various gravity models.

The condition $\frac{d\Theta_1}{d\tau}<0$ in the above equation
describes non-repulsive behavior of the gravitational interaction,
which can be written alternatively as $R_{\alpha\beta}k^\alpha
k^\beta\geq0$. In order to obtain this constraint, we have assumed
hypersurface orthogonal congruences to obtain
$\omega_{\alpha\beta}\equiv 0$ along with spatial shear tensor to be
$\sigma^2\equiv\sigma_{\alpha\beta}\sigma^{\alpha\beta}\geq0$. The
condition $R_{\alpha\beta}k^\alpha k^\beta\geq0$ can be written
instead as through energy-momentum tensor as stress-energy tensor
given by $T_{\alpha\beta}k^\alpha k^\beta\geq0$. For modified
theories of gravity (in term of effective density and pressure), the
ECs are
\begin{align}
\textrm{NEC}:&  {\rho _{eff}} + {p_{eff}} \ge 0,\\
\textrm{WEC}:& {\rho _{eff}} \ge 0 \text{ and } {\rho _{eff}} + {p_{eff}} \ge 0,\\
\textrm{SEC}:& {\rho _{eff}+ 3{p_{eff}}} \ge 0 \text{ and } {\rho _{eff}} + {p_{eff}} \ge 0,\\
\textrm{DEC}:&  {\rho _{eff}} \ge 0 \text{ and } {\rho _{eff}} \pm {p_{eff}} \ge 0.
\end{align}
Now, we write the cosmological parameters, i.e., deceleration, jerk, snap
and crackle, with the help of Hubble parameter H as
\begin{align}\label{ppvalues}
q =  - \frac{1}{{{H^2}}}\frac{{a''}}{a},\quad j = \frac{1}{{{H^3}}}\frac{{a'''}}{a},\quad
s = \frac{1}{{{H^4}}}\frac{{a''''}}{a},\quad
l =  - \frac{1}{{{H^5}}}\frac{{a'''''}}{a}.
\end{align}
In terms of these, the first, second, third and fourth differential forms of the
Hubble parameter are found as follows
\begin{align}\label{H1}
H' &=  - {H^2}\left( {1 + q} \right)\\\label{H2}
{H^{\prime \prime }} &= {H^3}\left( {j + 3q + 2} \right)\\\label{H3}
{H^{\prime \prime \prime }} &= {H^4} \left( {s - 2j - 5q - 3} \right)\\\label{H4}
{H^{{\prime \prime \prime \prime }}} &= {H^5}\left( {24 + l - 2j\left( { - 10 + q} \right)
+ 60q + 30{q^2} - 5s} \right).
\end{align}
Equations (\ref{tf1}) and (\ref{tf2}), after using
Eqs.(\ref{ppvalues})-(\ref{H4}) can be written for WEC as
\begin{align}\nonumber
\rho  &=  - \frac{f}{2} + f_{\Box R}^{\prime \prime \prime } + 3(2f_{\Box R}({H^{\prime \prime \prime }}
+ {H^4}\left( {14 + 3j + 23q + 6{q^2}} \right))+ H\\\label{ro}
&\times \{ { - {f_R}Hq - {H^2}\left( {5 + j + q} \right)f_{\Box R}^\prime
+ {f_R}^\prime  + H\left( {1 + 3q} \right)f_{\Box R}^{\prime \prime }} \}),\\\nonumber
\rho  + p &= {f_{\Box R}}^{\prime \prime \prime \prime } - 2{f_R}{H^2}
+ 112{f_{\Box R}}{H^4} + {f_\Box R}^{\prime \prime \prime }\left( {1 + 5H} \right)
+ 8{f_{\Box R}}{H^{\prime \prime \prime }}+ 24\\\nonumber
 &\times{f_{\Box R}}{H^4}j - 2{f_R}{H^2}q + 184{f_{\Box R}}{H^4}q
 + 48{f_{\Box R}}{H^4}{q^2}- 44{H^3}{f_{\Box R}}^\prime  - 7\\\label{ropp}
 &\times{H^3}j{f_{\Box R}}^\prime  - 13{H^3}q{f_{\Box R}}^\prime
 + H{f_R}^\prime  + 16{H^2}{f_{\Box R}}^{\prime \prime } + 14{H^2}
q{f_{\Box R}}^{\prime \prime } - {f_R}^{\prime \prime }.
\end{align}
In the above equations, we have assumed $f_T = 0$ in order to achieve simplicity.

\section{Different Models}

This section is devoted to examine some particular formulations of $f(R,\Box R,T)$
gravity for the FRW cosmic model on the behavior of the ECs. Ir order to
proceed forward our computation, we shall take the following specific cosmic
parameters
\begin{align}\label{pvalues}
H = 0.718,\quad q =  - 0.64,\quad j = 1.02,\quad s =  - 0.39,\quad l = 3.22.
\end{align}
In the following subsections, we model three different particular FRW models
with extra degrees of freedom coming from $f(R,\Box R,T)$ models.

\subsection{Model 1}

\begin{figure} \centering
\epsfig{file=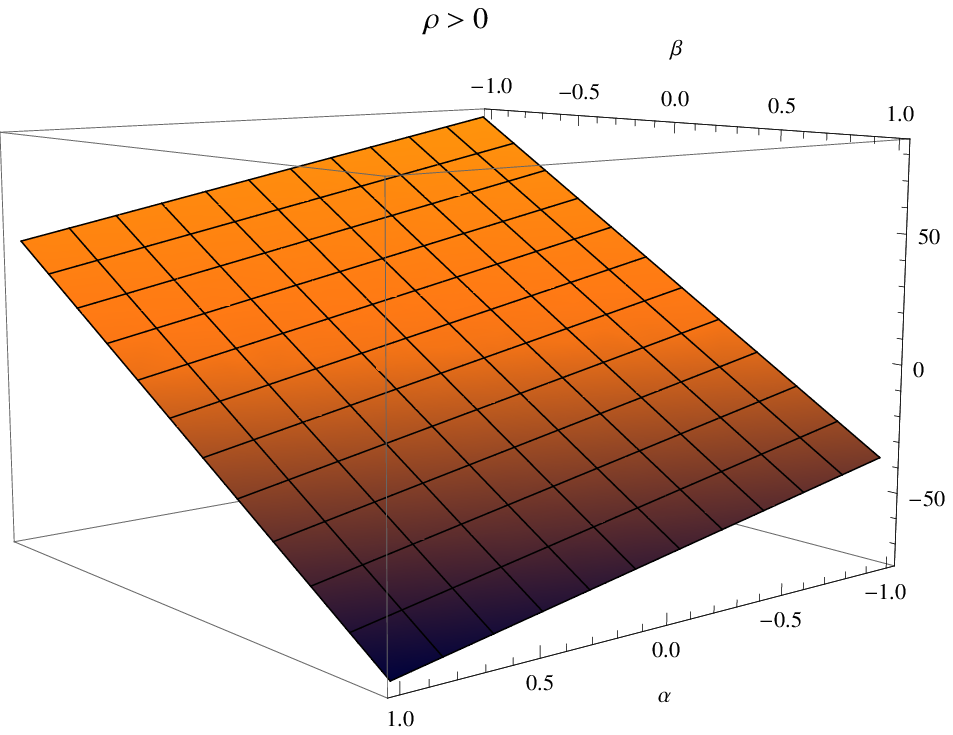,width=.48\linewidth}
\epsfig{file=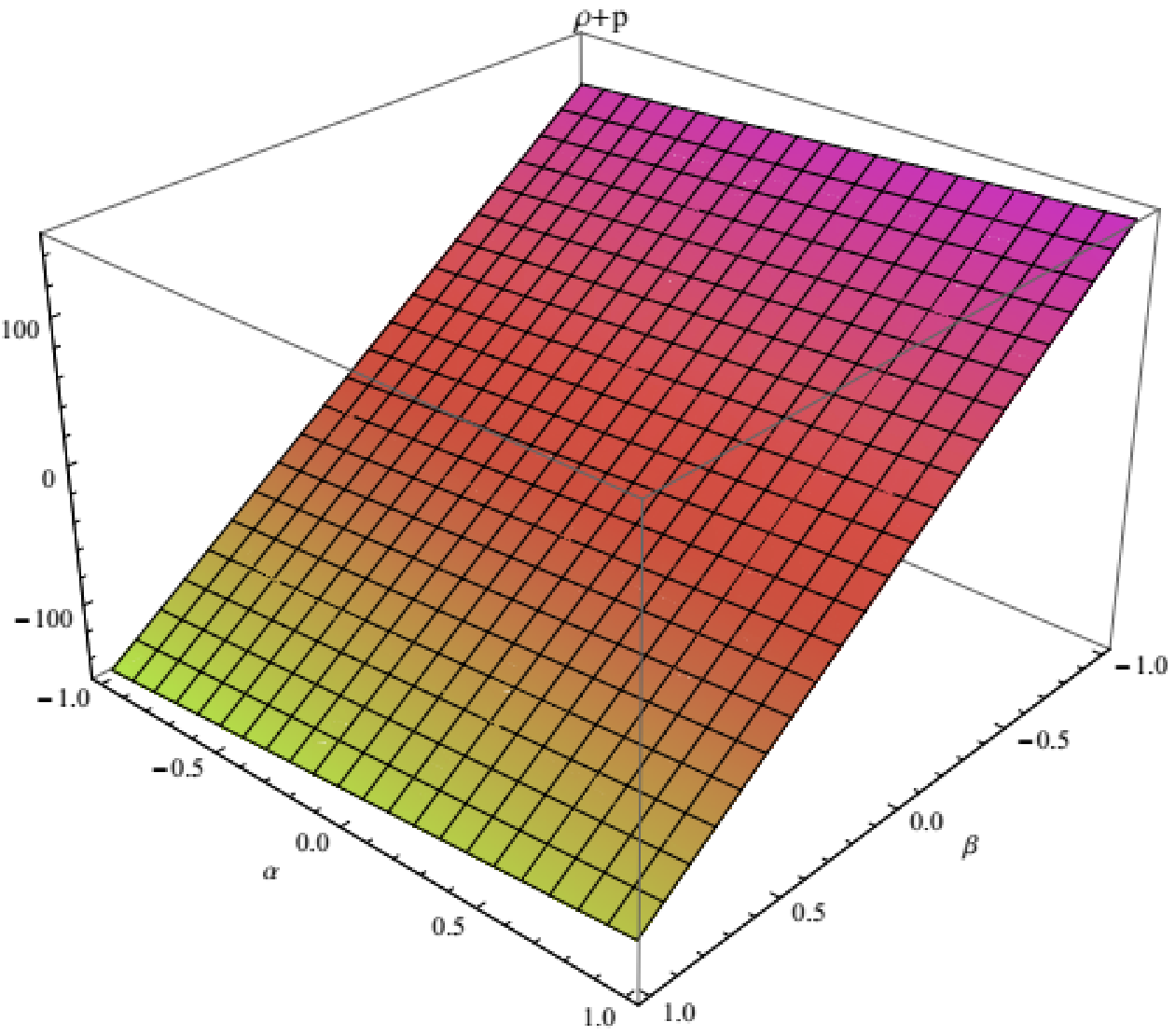,width=.48\linewidth} \caption{Plot of WEC for
model 1 as in Eq.(\ref{model1}), the left plot shows $\rho$ while
the right plot shows $\rho+ p$ with respect to $\alpha$ and
$\beta$}\label{f1}
\end{figure}

It would be very fascinating to add the higher derivative term in the quadratic Ricci scalar model of the form
\begin{equation}\label{model1}
f(R,\Box R) = R + \alpha {R^2} + \beta R \Box R,
\end{equation}
where $\alpha$ and $\beta$ are constant numbers. After using Eq.(\ref{pvalues}), Eqs.(\ref{ro}) and (\ref{ropp}) provide
\begin{align}\nonumber
\rho & = 3{H^2}(1 - 2q - 6\alpha {H^2}\left( { - 3 + 2j - 6q + 3{q^2}} \right)
+ 2\beta {H^3}(12 - l + 2\left( {10 + 7j} \right)q\\\nonumber
& + 6{q^2} + s) + 6\beta {H^4}( - 11 + {j^2} - l + 23q + 27{q^2} + 8{q^3} + j\left( {3 + 5q} \right) + qs)),
\end{align}
and
\begin{align}\nonumber
\rho  + p &= 2{H^2}[- 1 - q + 6\alpha {H^2}\left( {13 + j + 16q + 2{q^2} + s} \right)
+ 3\beta {H^3}(12 - l + 2(10 \\\nonumber
& + 7j )q + 6{q^2} + s) + 3\beta {H^4}( - 83 + 4{j^2} - 9l - 24q + 45{q^2} + 24{q^3} + j( 3 \\\nonumber
&+ 113q) - 4s + 2qs)].
\end{align}
We are concerned about examining the inequalities for the satisfaction of WEC
for the above-mentioned model. We have presented some graphs in
order to locate viability zones for the FRW perfect fluid model. The
validity regimes for WEC with the above model requires $\beta<0$. We
have shown through graphical representations (Figure \ref{f1}) that
viable universe models could possibly exist in zones, where space of
parameters allows negative values of $\beta$ without involving
exotic matter. In Figure \ref{f1}, the left plot describes $\rho>0$
and the right plot reports $\rho+p>0$.

Starobinsky \cite{u29} proposed that the quadratic Ricci scalar
corrections, i.e., $f(R)=R+\alpha R^2$ in the field equations could
be helpful to cause exponential early universe expansion. Various
relativists \cite{u30} adopted this formulation not only for an
inflationary constitute but also as a substitute for DM with
$\epsilon=\frac{1}{6M^2}$ \cite{u31}. For DM model, $M$ is figured
out as $2.7\times10^{-12}GeV$ with
$\epsilon\leq2.3\times10^{22}Ge/V^2$ \cite{u19a}. It is interesting
to mention that extension to this model could provide a platform
different from that of $R+\alpha R^2$ to understand various cosmic
puzzles even bounce cosmologies. The particular selection could be
constructive to examine inflation along with a stable background of the
scalar potential of an auxiliary field. This also helps to obtain a
potential having a non-zero residual vacuum energy, thereby
providing it as a DE in the late-time cosmic evolution. The
Starobinsky model is compatible with the recent joint investigation
of the B-mode polarization data from Keck/BICEP2 and Planck
temperature data and the array with the Planck maps at larger
frequencies. The tensor-to-scalar ratio is constrained to be
$r<0.08$ at $95\%$ confidence level. Ozkan {et al.} \cite{aa}
discussed the Planck constraints on inflation in auxiliary modified
$f(R)$ gravity. Odintsov and Oikonomou \cite{bb} provided a
qualitative comparison of non-singular version of the Starobinsky
$R^2$ inflationary model with the singular $R^2$ model. They claimed
that inflation ends more abruptly for singular $R^2$ mode as
compared to the ordinary inflationary model. Given these footing,
single field inflation establishes a scenario which is in full
agreement with the data. However, the nature of the inflaton is
still ambiguous.

\begin{figure} \centering
\epsfig{file=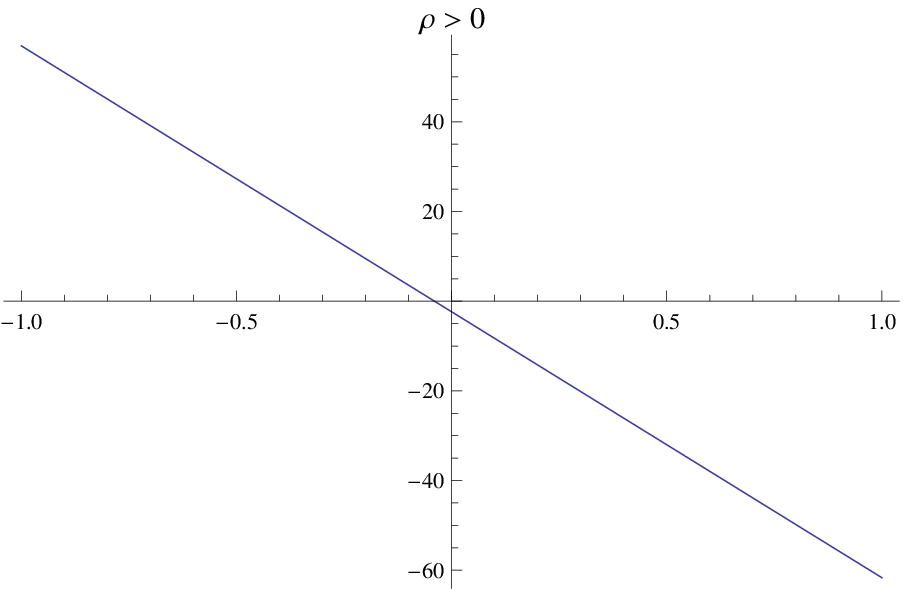,width=.48\linewidth}
\epsfig{file=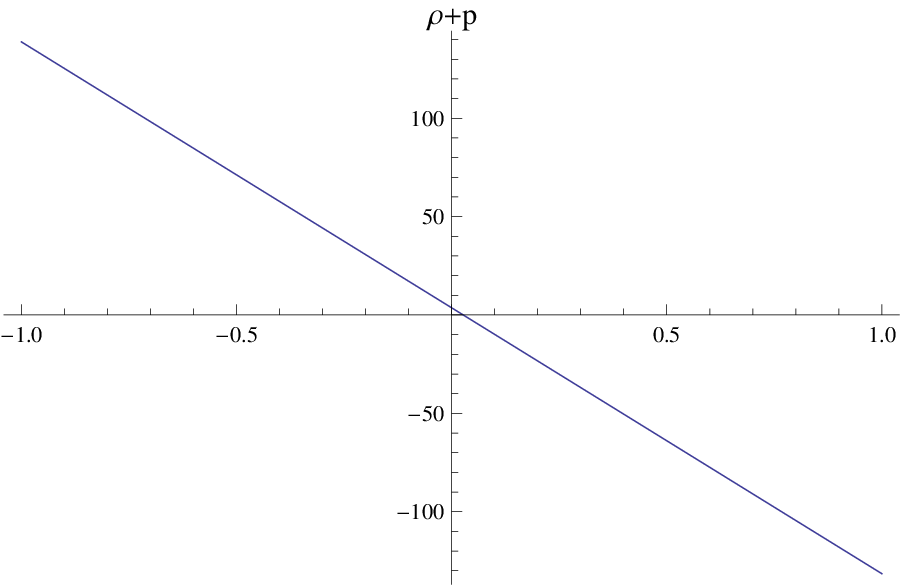,width=.48\linewidth} \caption{Contour plots of
WEC for model 1, the left plot shows $\rho$ while the right plot
shows $\rho+ p$ versus $\beta$.}\label{f1a}
\end{figure}

For our first model, we have two parameters, $\alpha$ and $\beta$
for which we examine the constraints on the parameters by plotting
$\rho$ and $\rho+p$ verses these parameters (as a result of 3D
plots). We modify Starobinsky model by adding an additional factor
($\Box R$) then by validity of energy condition, the values of these
parameters were fixed as indicated in Figure \ref{f1}. We
concluded that the WEC violates for positive values of both $\alpha$
and $\beta$. However, the WEC holds for $-1<\beta<0$ with small
positive values of $\alpha$. Using the constraints on the values of
$\alpha$, we have made the contour plots for both $\rho$ and
$\rho+p$. The range of $\alpha$ (chosen here) is compatible with the recent
Planck data for which the WEC holds. These results are shown in
Figure \ref{f1a}. The unit of $\alpha$ is the
same as that of $\frac{1}{M^2}$ while the unit of $\beta$ can easily
be evaluated through dimensional analysis.

\subsection{Model 2}

\begin{figure} \centering
\epsfig{file=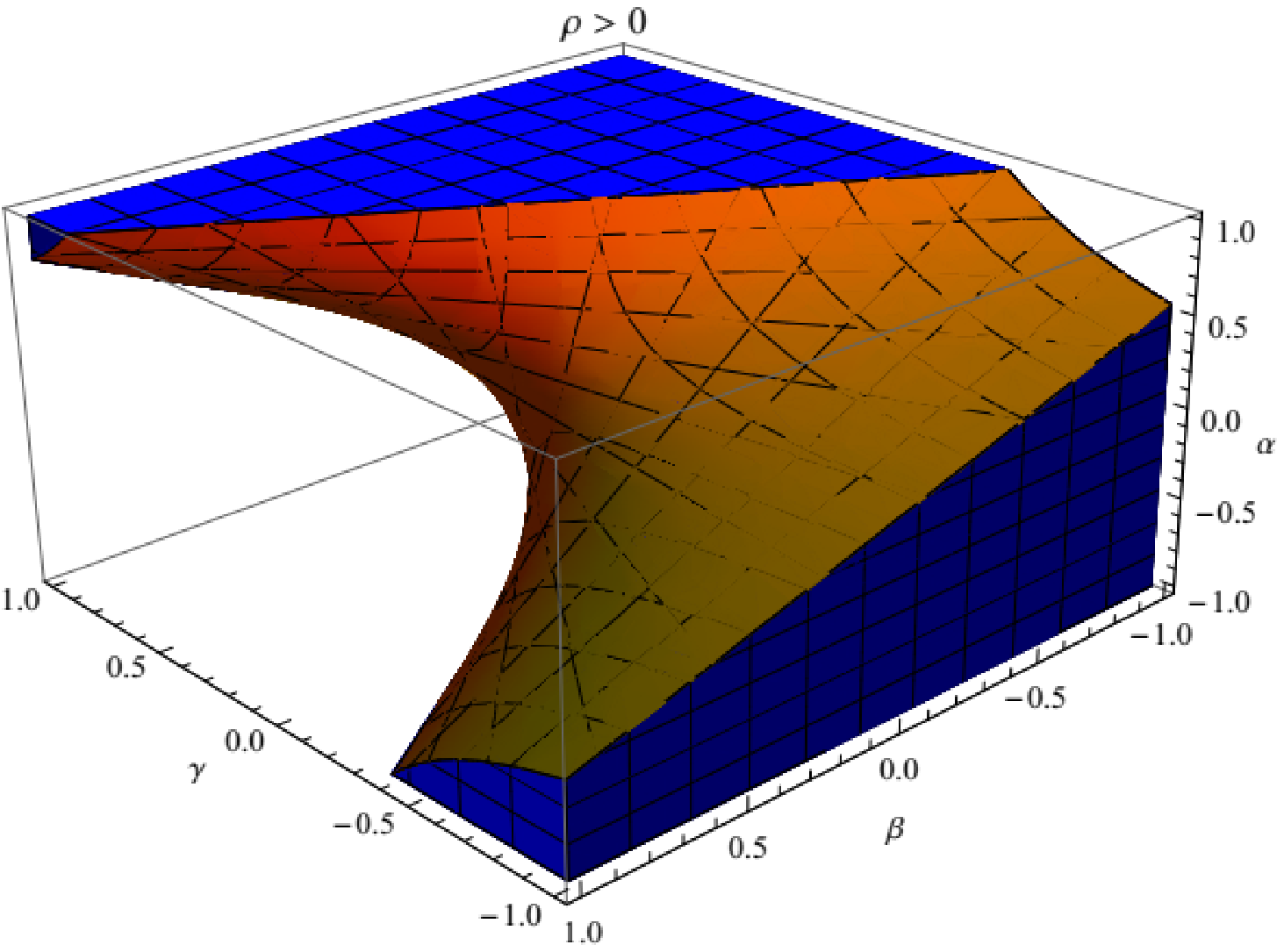,width=.48\linewidth}
\epsfig{file=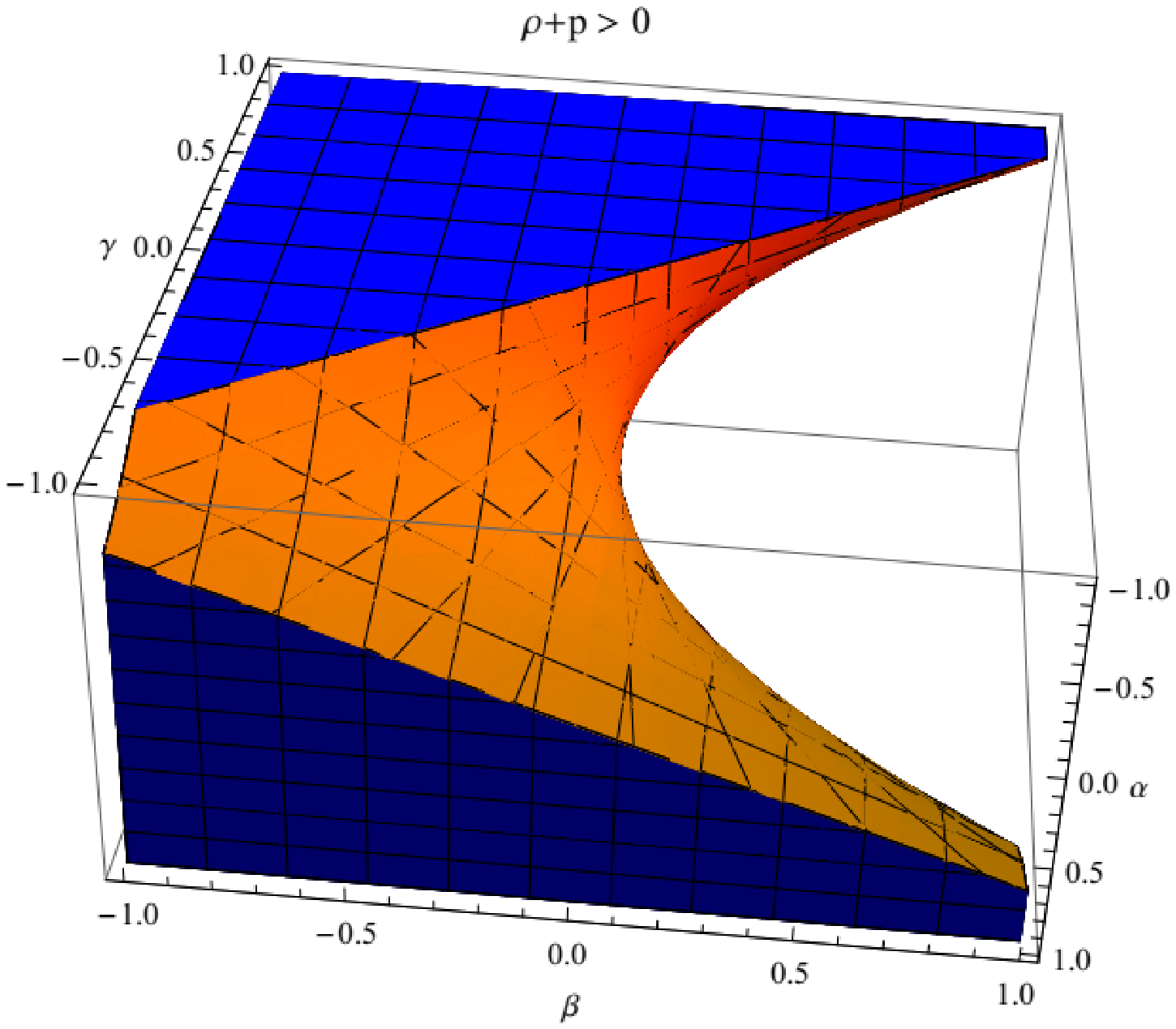,width=.48\linewidth} \caption{Plot of WEC for
model 2 as in Eq.(\ref{model2}), the left plot shows $\rho >0$ while
the right plot represents $\rho+ p >0$ with respect to $\alpha$,
$\beta$ and $\gamma$.} \label{f2}
\end{figure}

Next, we consider an interesting higher derivative model of the type
\begin{equation}\label{model2}
f(R,\Box R) = R + \alpha {R^2}\left( {1 + \gamma R} \right) + \beta R \Box R,
\end{equation}
where $\alpha,~\beta$ and $\gamma$ belong to set of real numbers.
Equations (\ref{ro}) and (\ref{ropp}), after using
Eqs.(\ref{pvalues}) and (\ref{model2}) with some manipulation, give
rise to
\begin{align}\nonumber
\rho& = 3{H^2}(1 - 2q - 6\alpha {H^2}( - 3 + 2j - 6q + 3{q^2}) + 2\beta {H^3}(12 - l
+ 2\left( {10 + 7j} \right)q \\\nonumber
& + 6{q^2} + s) + 6{H^4}( - 11\beta  - 66\alpha \gamma  + \beta {j^2}
- \beta l+ 23\beta q + 27\beta {q^2} + 90\alpha \gamma {q^2} + 8\beta \\\nonumber
&\times {q^3} - 24\alpha \gamma {q^3} + j\left( {3\beta  + 36\alpha \gamma
+ 5\beta q - 36\alpha \gamma q} \right) + \beta qs)),
\end{align}
while, it follows from the sum of energy density and pressure that
\begin{align}\nonumber
\rho+p  &= 2{H^2}( - 1 - q + 6\alpha {H^2}\left( {13 + j + 16q + 2{q^2} + s} \right)
+3\beta {H^3}(12 - l+ 2 \\\nonumber
&\times\left( {10 + 7j} \right)q + 6{q^2} + s) + 3{H^4}( - 83\beta  - 576\alpha \gamma
+4\left( {\beta  - 9\alpha \gamma } \right){j^2}
- 9\beta l\\\nonumber
& - 24\beta q - 288\alpha \gamma q + 45\beta {q^2} + 432\alpha \gamma {q^2}+24\beta {q^3}
+ 108\alpha \gamma {q^3} + j\{3(\beta + 36 \\\nonumber
&\times \alpha \gamma ) +(113\beta  + 108\alpha \gamma )q \}-4\beta s - 36\alpha \gamma s
+ 2\beta qs + 36\alpha \gamma qs)).
\end{align}
The ECs corresponding to the cosmic FRW model mediated by the above
mentioned model can be obtained through various numerical graphs. We
have explored these conditions and found that the inequalities of
WEC can be fulfilled by the perfect FRW metric under the following
parametric zonal values of spacetime. The validity region ($\rho>0$)
imposes that for $\beta$ to be in $[-1,1]$ makes the values of
$\gamma$ and $\alpha$ to be the open intervals $(-0.4 , -1)$ and
$(-0.4, 0.5)$, respectively. One can also find the regions where
$\rho>0$ by considering $\alpha$ belonging to $[-1,1]$ with
negative and positive values of $\beta$ and $\gamma$, respectively.
Similarly, the validity regions for $\rho+p >0$ have been
investigated in Figure \ref{f2}. It is observed that for all choices
of $\beta$ from $[-1,1]$ with $\gamma\in(-1,-0.7)$ and $\alpha\in
(-0.6, 1)$ give $\rho+p >0$. Another region has been seen for the
positivity of $\rho+p$ under which the tetrad of $\beta,~\gamma$ and
$\alpha$ should be from $[-1,1],~\gamma\in(0.6,1)$ and
$\alpha\in(-1, 0.7)$, respectively. We have shown graphically the
energy condition, as shown in Figure \ref{f2}, in which the left
plot is for $\rho>0$ while the right plot is for $\rho+p>0$.

\begin{figure} \centering
\epsfig{file=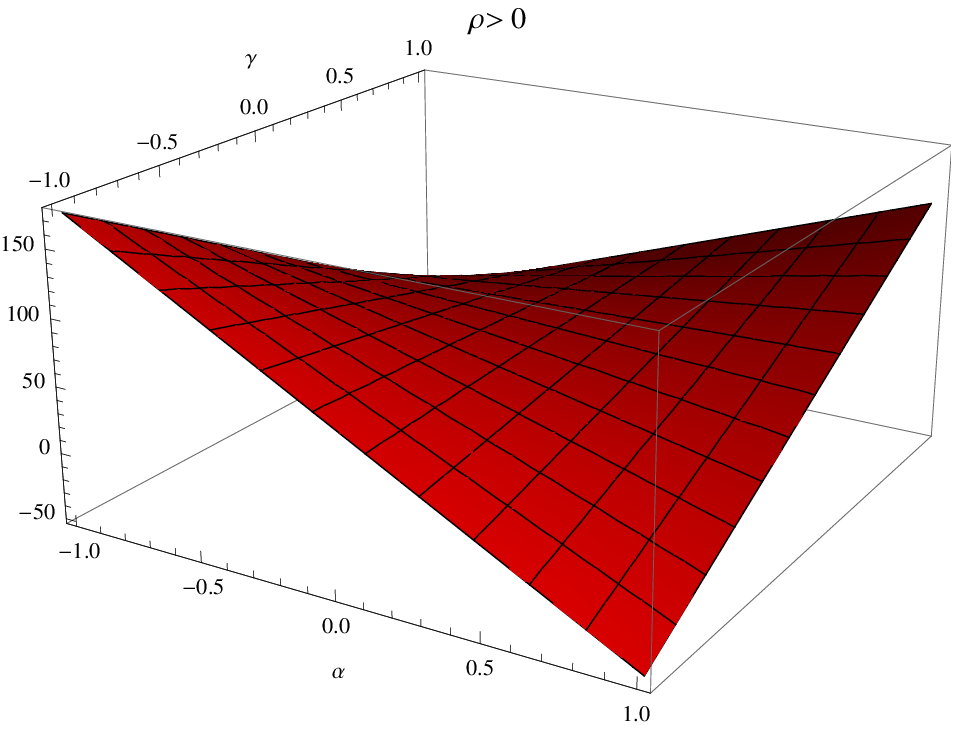,width=.48\linewidth}
\epsfig{file=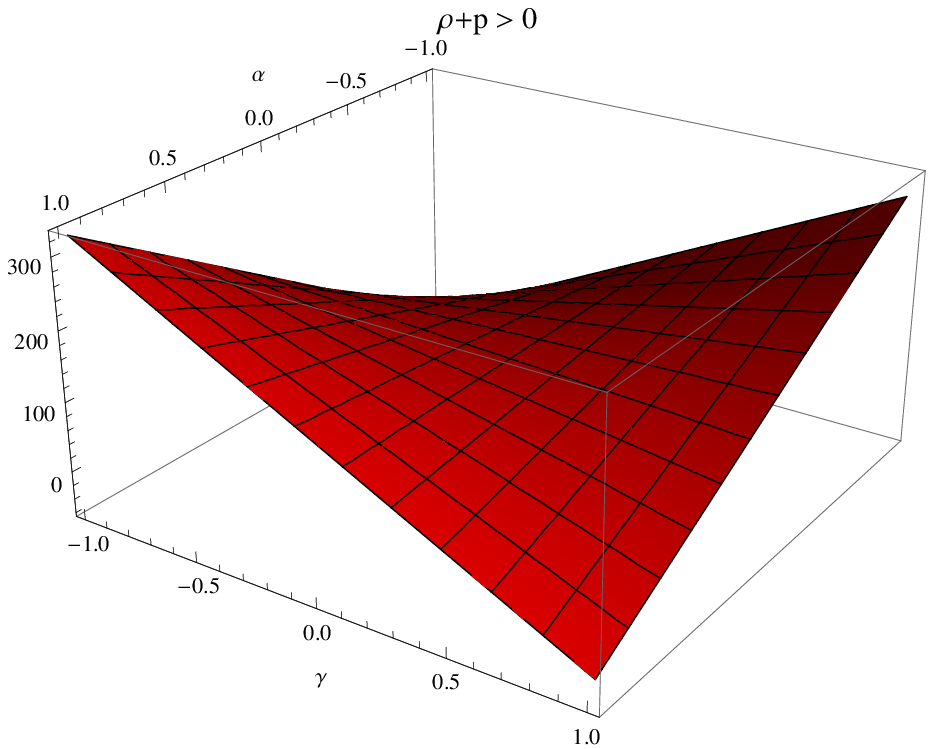,width=.48\linewidth} \caption{Plot of WEC
for model 2 with respect to $\alpha$ and $\gamma$.} \label{f2a}
\end{figure}

\begin{figure} \centering
\epsfig{file=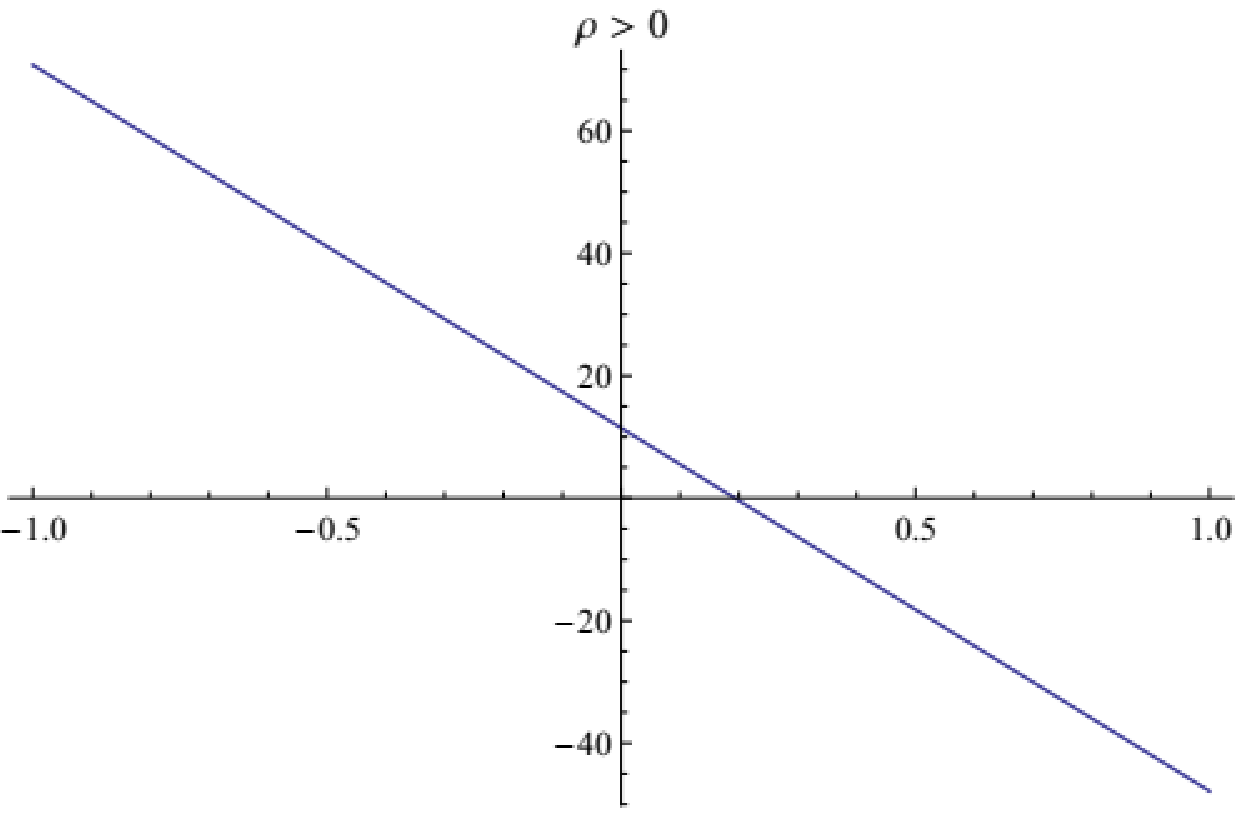,width=.48\linewidth}
\epsfig{file=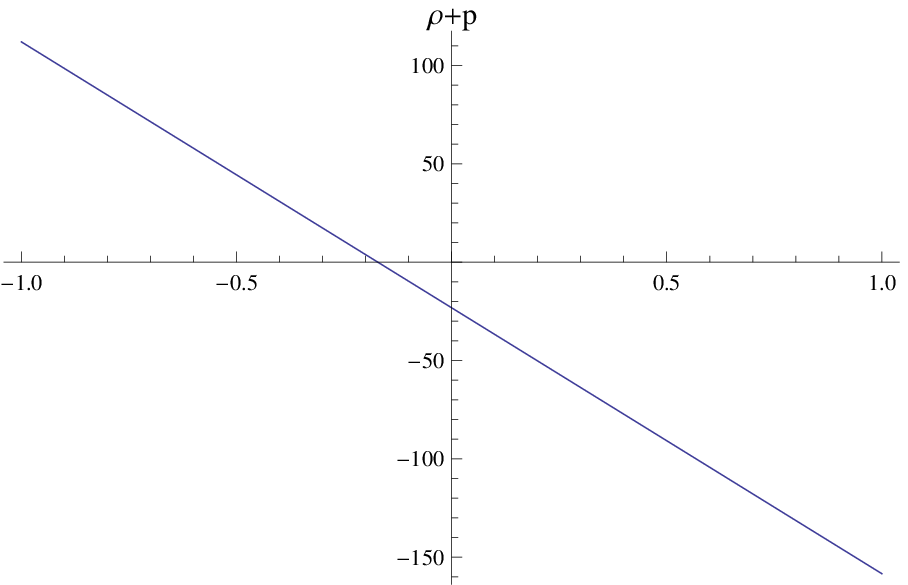,width=.48\linewidth} \caption{Plots of
WEC for model 2 versus $\beta$ for particular values of $\alpha$ and
$\gamma$.} \label{f2b}
\end{figure}

Model 2 could be treated as another modification
of Starobinsky inflationary model with an addition of higher
derivative term ($\Box$ R). In view of its compatibility with
latest Planck data, the parameter $\alpha$ restricts the parameter
$\beta$ and $\gamma$. We have explored that for which
values of $\alpha,~\beta$ and $\gamma$ the WEC holds. These are shown in the Figure \ref{f2}.
We have plotted the 3D graph to find the range of $\gamma$ for which
the WEC holds corresponding to $\beta=-1$. Consequently, we have
verified the range of $\beta$ for particular values of $\alpha$ and
$\gamma$. These results are indicated in Figures \ref{f2a} and \ref{f2b}.\\

\subsection{Model 3}

\begin{figure} \centering
\epsfig{file=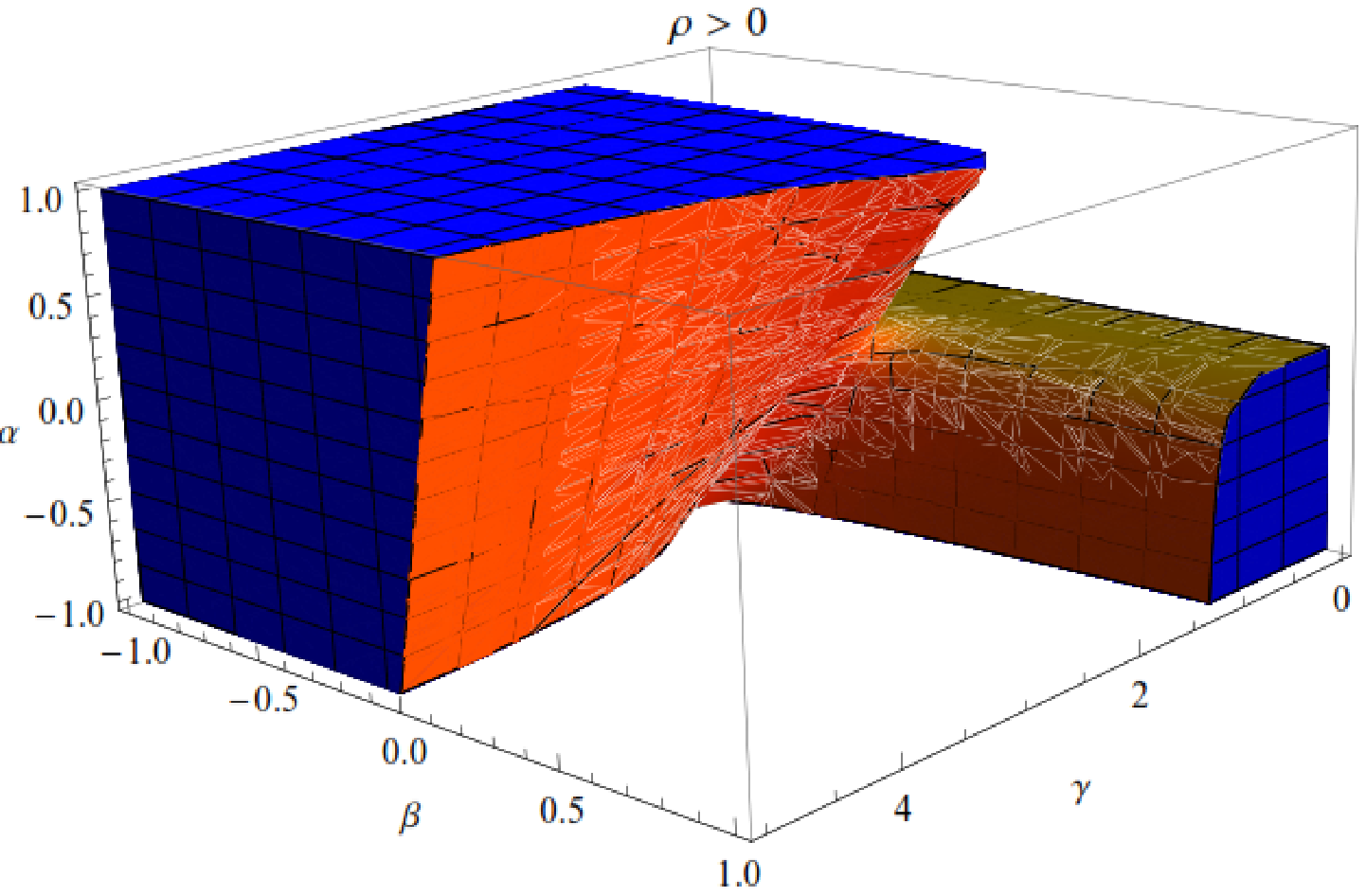,width=.48\linewidth}
\epsfig{file=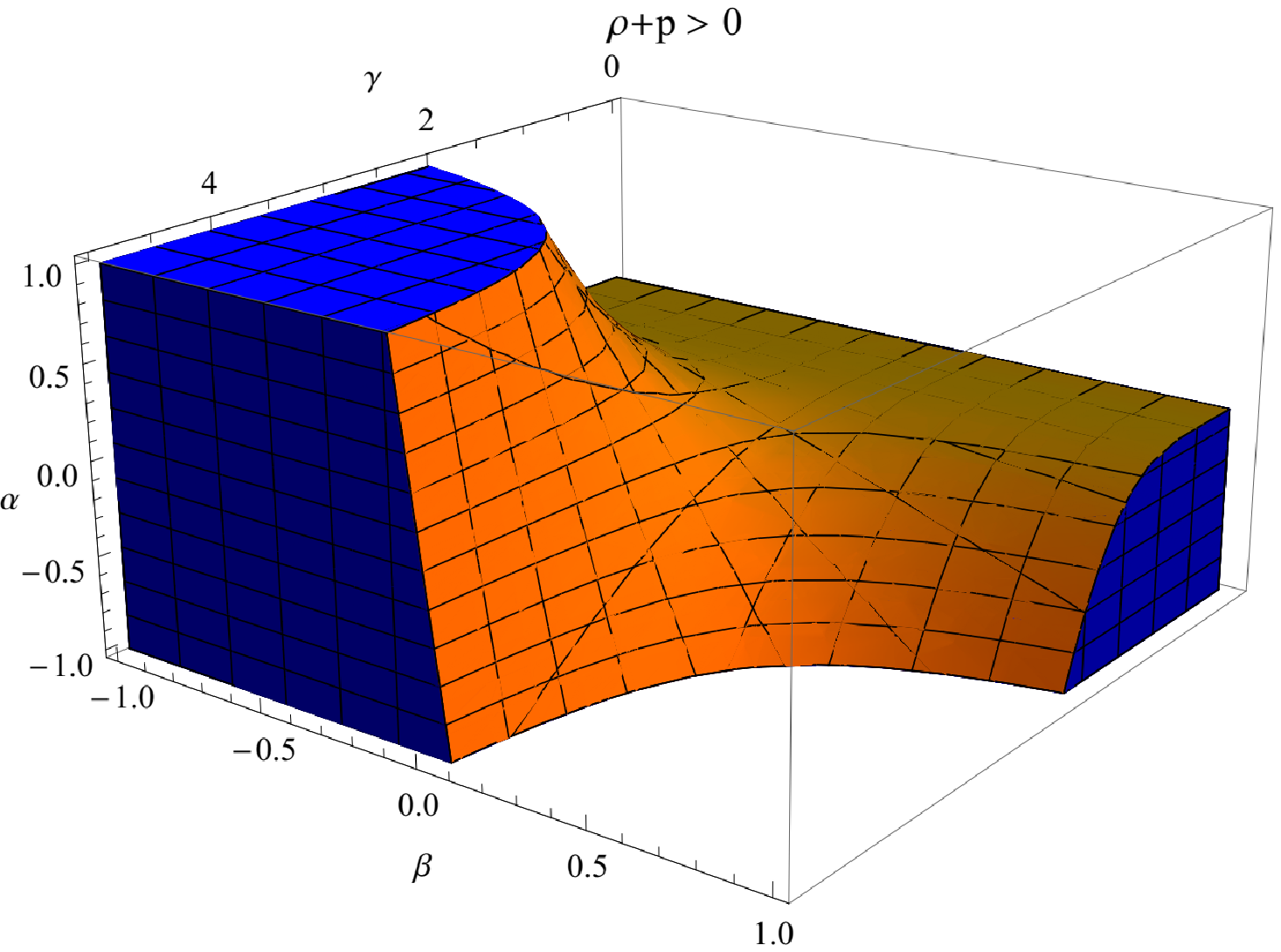,width=.48\linewidth} \caption{Plot of WEC for
model 3 as in Eq.(\ref{model3}), the left plot indicates $\rho
>0$ while the right plot shows $\rho+ p >0$ with respect to
$\alpha$, $\beta$ and $\gamma$.} \label{f3}
\end{figure}

Now, we consider modified curvature terms of the form
\begin{equation}\label{model3}
f = R + \alpha R\left[ {exp( - R/\gamma ) - 1} \right] + \beta R \Box R.
\end{equation}
This model could consistent DE results, which was introduced (with
$\beta=0$) in \cite{n1}. Furthermore, one can study details about
similar models on some $f(G)$ gravity which was first introduced in
\cite{n2}. By making use of expressions of snap, jerk and cackle
along with Eqs.(\ref{ro}) and (\ref{ropp}) , we get the following
form of the equations of motion
\begin{align}\nonumber
\rho&= \frac{3H^2}{{{\gamma ^2}}}{{\rm{e}}^{ - \frac{{6{H^2}\left(
{1 + q} \right)}} {\gamma }}}\left[\left(
{{{\rm{e}}^{\frac{{6{H^2}\left( {1 + q} \right)}}{\gamma }}} \left(
{ - 1 + \alpha } \right) - {{\rm{e}}^{\frac{{12{H^2}}}{\gamma
}}}\alpha } \right) {\gamma ^2}\left( { - 1 + 2q} \right)+
6{{\rm{e}}^{\frac{{12{H^2}}}{\gamma }}}\alpha \right.\\\nonumber
&\times\left. \gamma {H^2}( - 4 + 2j - 3q + {q^2})  +
2{{\rm{e}}^{\frac{{6{H^2} \left( {1 + q} \right)}}{\gamma }}}\beta
{\gamma ^2}{H^3}(12 - l + 2\left( {10 + 7j} \right)q + 6{q^2}\right.
\\\nonumber &\left.+s)+ 6{{\rm{e}}^{\frac{{6{H^2}}}{\gamma }}}{H^4}(
- 12{{\rm{e}}^{\frac{{6{H^2}}}{\gamma }}} \alpha -
11{{\rm{e}}^{\frac{{6{H^2}q}}{\gamma }}}\beta  {\gamma ^2} +
{{\rm{e}}^{\frac{{6{H^2}q}}{\gamma }}}\beta {\gamma ^2}{j^2} -
{{\rm{e}}^{\frac{{6{H^2}q}}{\gamma }}}\beta {\gamma ^2}l +
6{{\rm{e}}^{\frac{{6{H^2}}}{\gamma }}}\right. \\\nonumber &\times
\left.\alpha q+ 23{{\rm{e}}^{\frac{{6{H^2}q}}{\gamma }}}\beta
{\gamma ^2}q + 6{{\rm{e}}^{\frac{{6{H^2}}}{\gamma }}}\alpha {q^2} +
27{{\rm{e}}^{\frac{{6{H^2}q}}{\gamma }}}\beta {\gamma ^2}{q^2} +
8{{\rm{e}}^{\frac{{6{H^2}q}}{\gamma }}}\beta {\gamma ^2}{q^3} +
j(6{{\rm{e}}^{\frac{{6{H^2}}}{\gamma }}}\alpha\right. \\\nonumber &
\left.+ 3{{\rm{e}}^{\frac{{6{H^2}q}}{\gamma }}}\beta {\gamma ^2} +
\left( { - 6{{\rm{e}}^{\frac{{6{H^2}}}{\gamma }}}\alpha  +
5{{\rm{e}}^{\frac{{6{H^2}q}}{\gamma }}}\beta {\gamma ^2}} \right)q)
+ {{\rm{e}}^{\frac{{6{H^2}q}}{\gamma }}}\beta {\gamma ^2}qs)\right],
\end{align}
and
\begin{align}\nonumber
\rho+p  &= \frac{1}{{{\gamma ^3}}}2{{\rm{e}}^{ - \frac{{6{H^2}
\left( {1 + q} \right)}}{\gamma }}}{H^2}\left[\left(
{{{\rm{e}}^{\frac{{6{H^2} \left( {1 + q} \right)}}{\gamma }}}\left(
{ - 1 + \alpha } \right) - {{\rm{e}}^{\frac{{12{H^2}}}{\gamma
}}}\alpha } \right){\gamma ^3}\left( {1 + q} \right) +
108{{\rm{e}}^{\frac{{12{H^2}}}{\gamma }}}\right.\\\nonumber & \times
\left. \alpha {H^6}\left( { - 1 + q} \right) {\left( {2 - j + q}
\right)^2} - 6{{\rm{e}}^{\frac{{12{H^2}}} {\gamma }}}\alpha {\gamma
^2}{H^2}(12 + j+ 16q + 3{q^2} + s) + 3\right.\\\nonumber &
\times\left.{{\rm{e}}^{\frac{{6{H^2}\left( {1 + q} \right)}}{\gamma
}}}\beta {\gamma ^3}{H^3}(12 - l + 2\left( {10 + 7j} \right)q +
6{q^2} + s) + 3\gamma {H^4}\left\{\left( -
18{{\rm{e}}^{\frac{{12{H^2}}}
 {\gamma }}}\alpha\right.\right.\right.\\\nonumber
& \left.\left. \left. + 4{{\rm{e}}^{\frac{{6{H^2}\left( {1 + q} \right)}}{\gamma }}}\beta {\gamma ^2}\right){j^2}
+ j\left( 66{{\rm{e}}^{\frac{{12{H^2}}}{\gamma }}}\alpha
+ 3{{\rm{e}}^{\frac{{6{H^2}\left( {1 + q} \right)}}{\gamma }}}\beta {\gamma ^2}
+ \left( 42{{\rm{e}}^{\frac{{12{H^2}}}{\gamma }}}\alpha
+ 113\right.\right.\right.\right.\\\nonumber
&\left.\left.\left.\left.\times{{\rm{e}}^{\frac{{6{H^2}\left( {1 + q} \right)}}
{\gamma }}}\beta {\gamma ^2} \right)q \right) + {{\rm{e}}^{\frac{{6{H^2}}}{\gamma }}}
\left[ - 138{{\rm{e}}^{\frac{{6{H^2}}}{\gamma }}}\alpha  - 83{{\rm{e}}^{\frac{{6{H^2}q}}{\gamma }}}
\beta {\gamma ^2} - 9{{\rm{e}}^{\frac{{6{H^2}q}}{\gamma }}}\beta {\gamma ^2}l + 9 \right.\right.\right.\\\nonumber
& \times \left.\left.\left.(6{{\rm{e}}^{\frac{{6{H^2}}}{\gamma }}}
\alpha+ 5{{\rm{e}}^{\frac{{6{H^2}q}}{\gamma }}}\beta {\gamma ^2}){q^2}
+ 24\left( {{{\rm{e}}^{\frac{{6{H^2}}}{\gamma }}}\alpha  + {{\rm{e}}^{\frac{{6{H^2}q}}
{\gamma }}}\beta {\gamma ^2}} \right){q^3} - 6{{\rm{e}}^{\frac{{6{H^2}}}{\gamma }}}\alpha s - 4\right.\right.\right.\\
& \times\left.\left.\left.{{\rm{e}}^{\frac{{6{H^2}q}}{\gamma }}}\beta {\gamma ^2}s
+ 2q\left( {3{{\rm{e}}^{\frac{{6{H^2}}}{\gamma }}}\alpha s -
51{{\rm{e}}^{\frac{{6{H^2}}}{\gamma }}}\alpha  - 12{{\rm{e}}^{\frac{{6{H^2}q}}
{\gamma }}}\beta {\gamma ^2}  + {{\rm{e}}^{\frac{{6{H^2}q}}{\gamma }}}\beta {\gamma ^2}s} \right)\right\}\right]\right].
\end{align}
We want to check the validity regimes for WEC in this context. Through graphs,
the validity regions for $\rho>0$ impose the following constrains on the parameters of $f(R,\Box R, T)$ model\\
(i) For all $\alpha\in[-1,1]$, $\beta$ and $\gamma$ depend on each other, like $\beta < 0$ while $\gamma>2$.\\
(ii) For all $\beta\in[-1,1]$, $\alpha$ and $\gamma$ depend on each other, e.g $\alpha < 0$ then $\gamma <2$.\\
Similarly, the condition $\rho+p >0$ provides the following limitations on $\alpha,~\beta$ and $\gamma$\\
(i) For all $\alpha\in[-1,1]$, $\beta$ and $\gamma$ depend on each other, like $\beta < 0$ while $\gamma>2.1$.\\
(ii) For all $\beta\in[-1,1]$, $\alpha$ and $\gamma$ depend on each other, e.g $\alpha < 0$ then $\gamma <2.1$.\\
We have shown graphically the viable zones of WEC in Figure
\ref{f3}. It can be noticed that the left plot of the Figure
(\ref{f3}) describes $\rho>0$, its right plot provides zones for
$\rho+p>0$.

Modified gravity theories has a significant role in various forms in
the discussion of the universe's evolution. It is worth mentioning
that a number of viable $f(R)$ gravity models leading to a unified
description of inflation with late time cosmic acceleration have
been identified in literature. Elizalde {et al.} \cite{cc} proposed
a reasonably simple and a viable version of $f(R)$ gravity via an
exponential modified gravity model which explains in a unified way
both the late time acceleration and early time inflation and
successfully fulfils the cosmological bounds as well as compatible
with the known local tests. This exponential gravity is indeed free
from any kind of finite time future singularity and exhibit various
interesting properties. They also claimed that slight generalization
to this theory may yield other well-behaved non-singular stable
class of exponential gravity models with similar physical
predictions.  Bamba {et al.} \cite{dd} presented a detailed analysis
on the evolution history of the growth index of matter density
perturbation as well as the future evolution of the universe via the
exponential gravity model. They exhibited the numerical analysis of
inflation via couple of viable exponential gravity models. They also
investigated the behavior of extra correction ingredients in the
exponential gravity model to examine the cosmic acceleration and
inflation.

\begin{figure} \centering
\epsfig{file=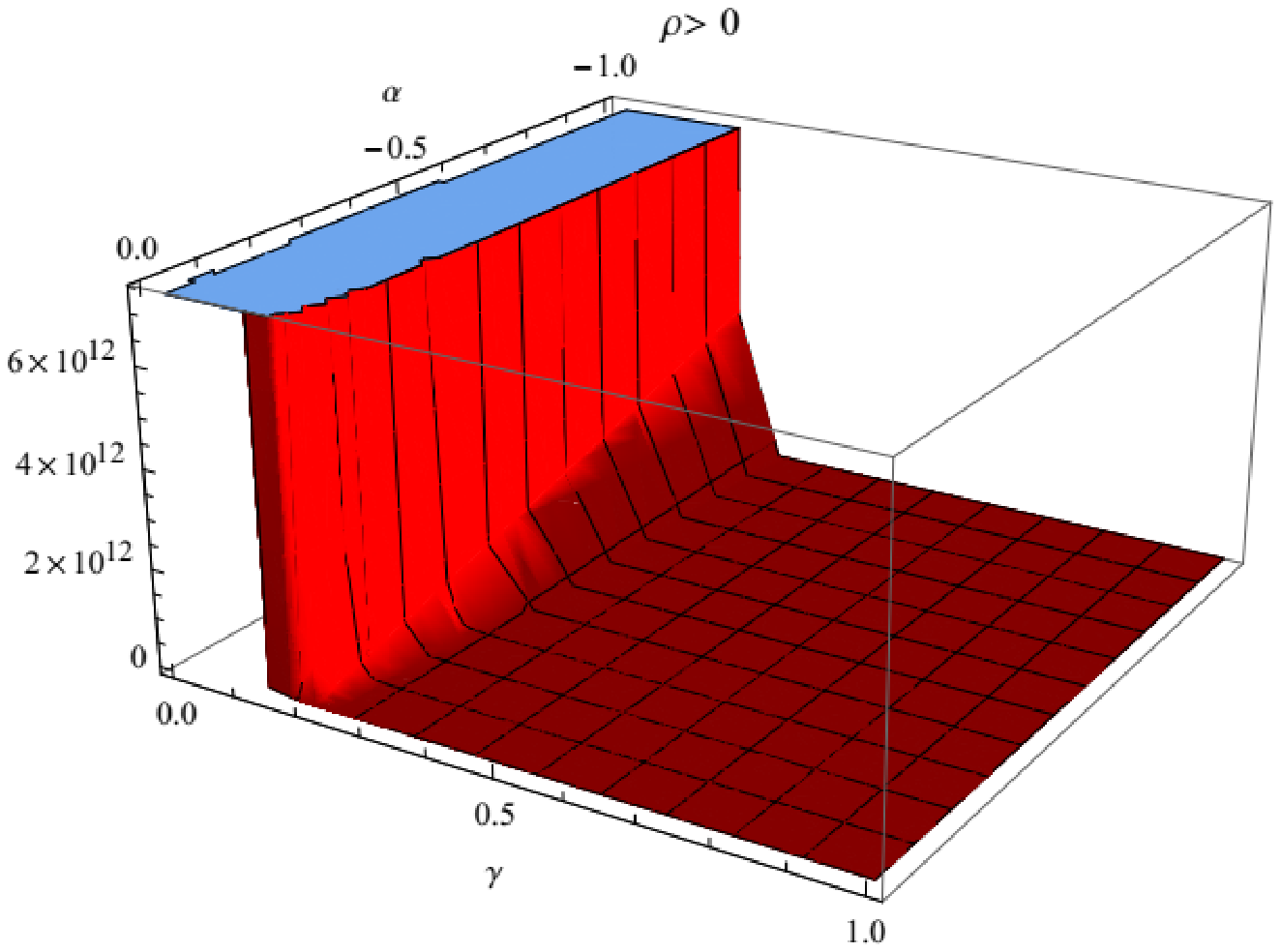,width=.48\linewidth}
\epsfig{file=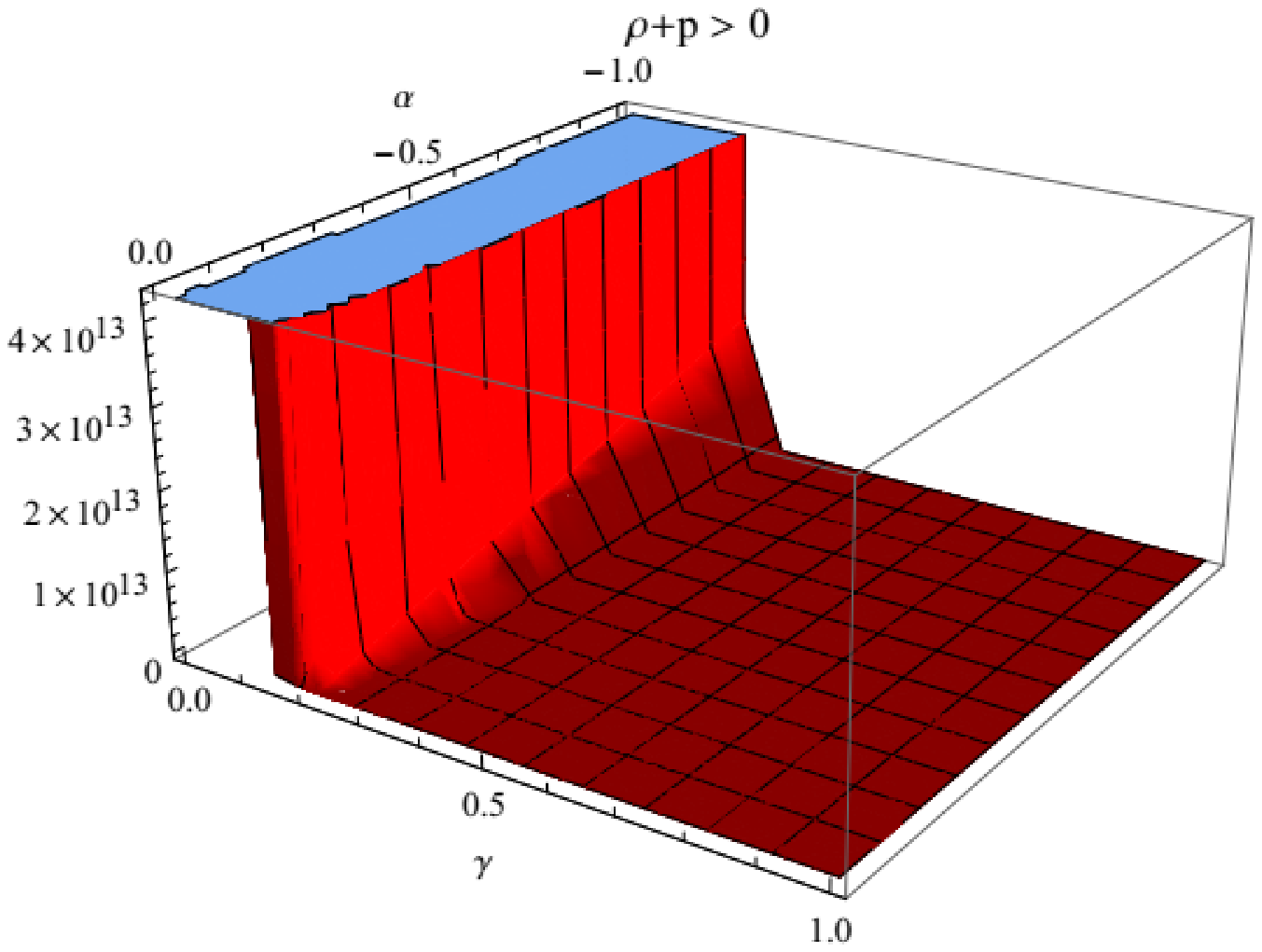,width=.48\linewidth} \caption{Plot of WEC
for model 3, the left plot indicates $\rho
>0$ while the right plot shows $\rho+ p >0$ with respect to
$\alpha$ and $\gamma$.} \label{f3a}
\end{figure}

\begin{figure} \centering
\epsfig{file=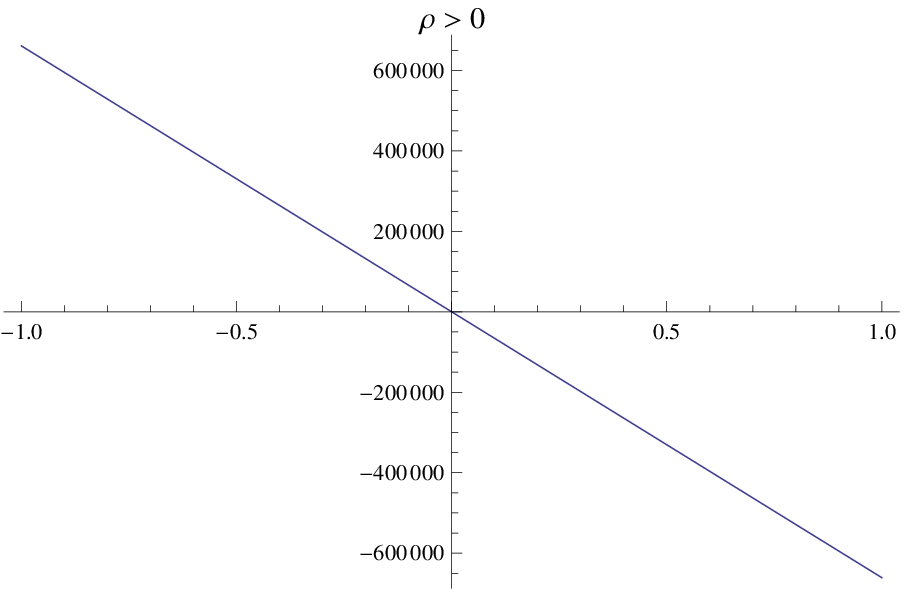,width=.48\linewidth}
\epsfig{file=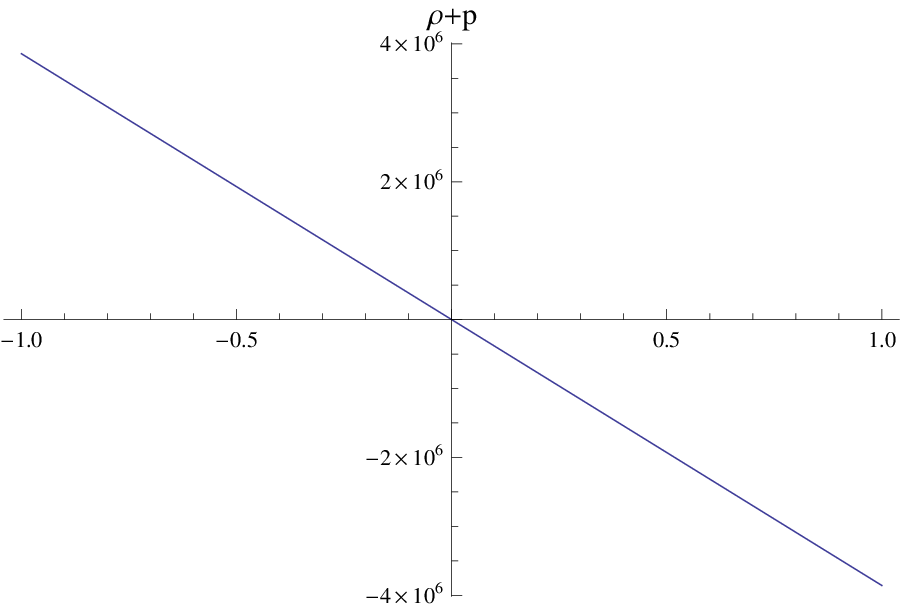,width=.48\linewidth} \caption{Plot
of WEC for model 3, the left plot indicates $\rho
>0$ while the right plot shows $\rho+ p >0$ for particular values of $\beta$
and $\gamma$ with respect to $\alpha$.} \label{f3b}
\end{figure}

Odintsov {et al.} \cite{ee} provided various interesting properties
of the reheating regime via the logarithmic correction in the $R^2$
gravity model as this provide a unification of late and early
acceleration era. They also discussed the possibility to explore
observational indices compatible with the latest Planck data. We
have considered an exponential gravity model to explore the
viability regions for WEC. In this case, we have plotted 3D the
results of WEC with negative values of $\beta$ to explore the range
of $\gamma$ where WEC holds (shown in Figure \ref{f3}). Further, we have plotted the WEC for
the extracted value of $\gamma$ and $\beta$ to examine the viability
regions for WEC. All this analysis have been shown via plots in
Figures \ref{f3a} and \ref{f3b}.

\section{Summary}

\vspace{0.5cm}

The existence of a generic higher-derivative curvature quantities
term in the extended version of EH action could help to provide some
dynamical features of canonical Einstein gravity (after using
conformal transformations) along with finite scalar degrees of
freedom. From these, one can notice that few fields mediated by
scalar variables are propagating with proper mathematical and
physical grounds. The present paper is devoted to investigate the
impact of $f(R,\Box R, T)$ extra curvature terms in the modeling of
realistic cosmic ideal fluid models. The Lagrangian of $f(R,\Box R,
T)$ gravity could be regarded as more comprehensive in its meaning
that suggests that various functional formulations of $f$ could be
introduced. Due to the arrival of a wide variety of  these
functions, the need of the hour is to constrain such gravitational
theory on mathematical as well as physical backgrounds. The validity
of this gravitational model could be dealt by analyzing the
stability constraints against local perturbations which is widely
known as Dolgov-Kawasaki instability.

This work is devoted to develop some conditions on comprehensive as
well as particular configurations of $f(R,\Box R, T)$ models by
investigating the applicability of respective ECs. We have used
expressions for NEC and SEC that could be calculated after using the
notable Raychaudhuri equation with an environment of the attractive
nature of gravity. This gives rise to even more generic computations
of results as compared to $f(R)$ and $f(R,T)$ gravitational models.
Furthermore, these expressions are found to be equivalent to that
determined by $\rho +3p\geq0$ and $\rho+p\geq0$ after using
particular mathematical transformations on mater variables, i.e.,
$p\rightarrow p_{eff}$ and $\rho\rightarrow \rho_{eff}$ ,
respectively. In a similar fashion, one can utilize the
corresponding constraints for WEC and DEC by transforming the
counterpart matter quantities in GR for $f(R,\Box R, T)$
energy-momentum tensor.

In order to evaluate the viability constraints for $f(R,\Box R, T)$
gravity, we have considered three particular cosmic models
specifically, $f(R,\Box R) = R + \alpha {R^2} + \beta R \Box
R,~f(R,\Box R) = R + \alpha {R^2}\left( {1 + \gamma R} \right) +
\beta R \Box R$ and $f(R,\Box R) = R + \alpha R\left[ {exp( -
R/\gamma ) - 1} \right] + \beta R \Box R$. We have used the
expressions of snap, jerk and crackle that are being evaluated
through Hubble parameter. It is shown that WEC for these models
depends on the particular choices of $\alpha,~\beta$ and $\gamma$.
The ECs are valid in the first model, once we set $\beta<0$. The
plot has been shown in order to mention the viability bounds of WEC
within the background of $f(R,\Box R) = R + \alpha {R^2} + \beta R
\Box R$ in Figure \ref{f1}. For the case of second model, i.e.,
$f(R,\Box R) = R + \alpha {R^2}\left( {1 + \gamma R} \right) + \beta
R \Box R$, we proceed our examination to obtain validity regions of
WEC. We found that zone satisfying $\rho+p>0$ constraint for the FRW
perfect model can be obtained by keeping $\beta$ to be in $[-1,1]$
along with $-1 < \gamma < -0.7$ and $-0.6 < \alpha < 1$.
Furthermore, the validity regions for $\rho+p>0$ can be found by
taking any choice of $\beta$ from $[-1,1]$ together with
$\gamma\in(0.6, 1)$ and $\alpha\in(-1,0.7)$. The viability bounds on
the formulations of second models for $\rho>0$ are found as
$\beta\in[-1,1]$ along with $\gamma\in (-1, -0.4)$ and $\alpha\in
(-0.4, 0.5)$. Figure \ref{f2} also indicates that conditions on the
parameters of $f$ model for the applicability of $\rho>0$, are
$\alpha\in[-1,1],~\beta < 0$ and $\gamma >
 0$. The EC $\rho>0$ for
$f(R,\Box R) = R + \alpha R\left[{\exp( - R/\gamma ) - 1} \right] +
\beta R \Box R$ is met, if we take $\alpha\in[-1,1]$ with negative
choice of $\beta$ and $\gamma$ to be greater than 2. Furthermore, to
keep $\beta\in[-1,1],~\alpha < 0$ and $\gamma <2$, the viability
zones for $\rho>0$ can be found as shown in Figure \ref{f3}. We
further found that one of the conditions of WEC, $\rho+p >0$, is
valid under the parametric limits $\alpha\in[-1,1]$ along with
$\beta < 0$ and $\gamma>2.1$. Also, the region governed by
$\beta\in[-1,1],~\alpha < 0$ and $\gamma <2.1$ provides the
viability bounds for $\rho+p >0$.

The terminology of ``modified gravity" has received great attention
that depicts gravitational interactions, differ from the most
established theory of general relativity. It is pertinent to mention here that the investigation for the
well-consistent $f(R,\Box R, T)$ gravity models can be extended for
the notion that there could exist some convenient usual complex
fluid configurations corresponding to FLRW geometry. Then, the
perspective research may give rise to some substantial qualitative
consequences in comparison with the discussion of pure gravity. It
will be executed elsewhere.

\end{document}